\documentstyle[12pt]{article}

\def\tend{\mathop{\to}}

\def\lim{\mathop{\rm {lim}}}

\begin{document}
\large
\rm
\begin{center}
\bigskip
{\bf  NONLOCAL INTERACTIONS AND QUANTUM DYNAMICS }\\
\bigskip
{\rm
Renat Kh.Gainutdinov}\\\bigskip
\it
Department of Physics\\
Kazan State University,\\
18 Kremlevskaya St, Kazan 420008,\\
Russia\\
E-mail: Renat.Gainutdinov@ksu.ru
\end{center}
\begin{center}
\bigskip
{\bigskip\bf Abstract}
\end{center}
\vskip 1.5em

\parindent=2.5em

{\rm\par

The problem is considered of describing the dynamics of quantum 
systems generated by a nonlocal in time interaction. 
It is shown that the use of the Feynman  approach to 
quantum theory in combination with the canonical approach allows 
one to extend quantum dynamics to describe the time 
evolution in the case of such interactions. 
In this way, using only the current concepts of 
quantum theory,  a generalized equation of motion for state 
vectors is derived. 
In the case, where the fundamental 
interaction generating the  dynamics 
in a system is local in time, this equation is equivalent to the 
Schr{\"o}dinger equation. 
Explicit examples are given for an exactly solvable model. 
The proposed formalism is shown to provide a new 
insight into the problem of the description of 
nonlocal interactions in quantum field theory. 
It is shown that such a property of the equation of 
motion as nonlocality 
in time may be important for describing hadron-hadron 
interactions at low and intermediate energies.

\section*{I. Introduction}

Various physical applications of quantum mechanics and quantum 
field theory require solutions of evolution problems. 
In standard canonical quantum theory, 
it is postulated that quantum  dynamics is governed 
by the  Schr{\"o}dinger equation. 
However, as is well known, in quantum electrodynamics 
(QED) the  ultraviolet divergences can be removed from the 
S matrix, but cannot be removed from quantities 
characterising the time evolution of quantum systems, 
since regularisation of the scattering matrix leads 
to a situation in which divergent terms 
automatically appear in the  Schr{\"o}dinger equation [1]. 
For this reason this equation is of only formal 
importance to quantum field theory. 
This leads, in particular, to difficulties in finding a 
consistent QED description of natural broadening 
of spectral line profiles in atomic system [2]. 
Since locality has been argued to be the 
main cause of infinities in quantum field theory, 
it seems natural to resolve this problem by introducing a 
nonlocal form factor into the interaction Hamiltonian density. 
However, as is well known, such an introduction of a nonlocal 
form factor results in a loss of covariance. 
The reason for this is quite obvious. 
The Schr{\"o}dinger equation is local in time, and 
the interaction Hamiltonian describes an instantaneous 
interaction.
In nonrelativistic quantum mechanics processes of 
instantaneous interaction may be nonlocal in space. 
But in relativistic quantum theory a local in time 
process must be also local in space. 
Thus, for the introduction of nonlocality in the theory 
to be intrinsic consistent, one has to find a way of 
solving the evolution problem in the case when the 
dynamics in a system is generated by a nonlocal in 
time interaction. 
The solution of this problem may be important also for 
describing the low-energy hadron-hadron interaction. 
Indeed, nonlocality of interaction leads to the energy 
dependence of effective interaction operators. 
In recent years the possibility of using energy-dependent 
potentials to describe hadron-hadron interaction at 
low and intermediate energies has been widely 
discussed [3-10]. 
Interest in studying potentials of this type is provoked 
by still existing discrepancy between theory and experiment. 
For example, most "realistic" nucleon-nucleon (NN) 
potentials are not sufficiently strong to reproduce 
the observed  ${}^3H$ and ${}^3He$ binding energies 
(see, for example, Ref.[11]). 
Moreover, the energy dependence of effective operators 
of hadron-hadron interaction is associated with the quark 
degrees of freedom, which are not included explicitly 
in the description of low-energy hadron interaction, and is 
predicted, for example, by quark models [3,10]. 
However, the dynamics of such systems cannot be described 
in the framework of the Hamiltonian formalism. 
Indeed, the energy dependence of the interaction Hamiltonian 
means that the total Hamiltonian also depends on energy, 
i.e. on its spectral parameter. 
Such an "operator" is not an operator in a rigorous sense. 
Hence the energy-dependent interaction operator cannot 
be interpreted as an interaction Hamiltonian. 

Let us now turn to the Feynman formulation of quantum 
theory [12,13]. 
The main idea of this formulation is that  quantum 
dynamics can be described without resorting to the Schr{\"o}dinger 
equation. 
Feynman's theory starts with an analysis of the phenomenon of 
quantum interference which leads directly to the concept 
of the superposition of probability amplitudes. 
According to this concept, the probability amplitude of an event 
which can happen in several different ways is a sum of 
probability amplitudes  for each of these way [12]. 
The Feynman formulation contains also as its essential 
idea the concept of a probability amplitude associated with a 
completely specified motion or path in space-time, 
and it is postulated that this probability amplitude has a 
phase proportional to the action, computed classically, for 
the corresponding path. 
Using this postulate together with the above assumption 
concerning the calculation of probabilities in quantum mechanics 
leads to Feynman's sum-over-paths formalism. 

The theory of Feynman differs profoundly in its formulation 
from canonical quantum theory. 
These dissimilar approaches were proved to be equivalent 
and to complement one another in solving various 
problems in quantum physics. 
In the present paper we show that these two approaches can be 
used in combination for describing the time evolution of 
quantum systems. 
In this way an equation of motion for state vectors 
is derived. 
Being equivalent to the Schr{\"o}dinger equation in the 
case of local interaction, this equation makes it 
possible to solve the evolution problem in the 
case where the interaction generating the dynamics 
in a quantum system is nonlocal in time. 

We start with the evolution equation $|\psi(t)>= U(t,t_0)|\psi(t_0)>,$ 
where $|\psi(t)>$ is a state vector and $U(t,t_0)$ is the 
evolution operator. 
Then, using the basic assumption of the Feynman formulation, we 
represent the matrix elements of the evolution operator 
as a sum of contributions from all alternative ways of realisation 
of the corresponding evolution process. 
The history of a quantum system is represented by 
some version of the time evolution of the system associated 
with completely specified instants of the beginning and end of the 
interaction in the system. 
In this way we get an expression for the matrix elements of the 
operator $U(t,t_0)$ in terms of probability amplitudes associated 
with such versions. 
An equation for these amplitudes is then derived from the requirement 
of unitarity for the evolution operator. 
It is shown that this equation can be regarded as an 
equation of motion. 
The concept of a generalized interaction operator is introduced. 
This operator is a generalization of the interaction Hamiltonian 
and generates the dynamics of a quantum system. 
In particular, the generalized interaction operator can be 
chosen so that the dynamics generated by this operator 
proves equivalent to the dynamics governed by the  
Schr{\"o}dinger equation. 
At the same time, the equation of motion derived in this 
paper permits the generalization to  
the case  where this equation manifests itself 
as a nonlocal in time dynamical  equation. 
This  point is illustrated in detail on an exactly 
solvable model. 
Finally, we discuss applications of the proposed formalism. 
This formalism is shown to provide a new insight into the 
problem of the description of nonlocal interactions in quantum 
field theory. 
It is shown also that such a property of the proposed dynamical equation 
as nonlocality in time may be important for describing 
hadron-hadron interactions at low and intermediate energies.

\section*{II. Basic Assumptions Of Quantum Theory}

We will assume the following properties for the states of
quantum systems.

(i) The physical state of a system is represented by a vector
(properly by a ray) of a Hilbert space.

(ii) An observable A is represented by a Hermitian hypermaximal operator
$\alpha$. The eigenvalues $a_r$ of $\alpha$ give the possible values of A.
An eigenvector $|\varphi_r^{(s)}>$ corresponding to the eigenvalue
$a_r$ represents a state in which A has the value $a_r$. If the system
is in the state $|\psi>,$ the probability $P_r$ of finding the value
$a_r$ for A, when a measurement is performed, is given by
$$P_{r} = <\psi|P_{V_{r}} |\psi>= \sum_s |<\varphi_r^{(s)}|\psi>|^2, $$
where $P_{V_{r}}$ is the projection operator on the eigenmanifold $V_r$
corresponding to $a_r,$ and the sum $\Sigma_s$ is taken over a complete
orthonormal set ${|\varphi_r^{(s)}>}$ (s=1,2,...) of $V_r.$
The state of the system immediately after the observation is
described by the vector $P_{V_{r}}|\psi>.$

These  assumptions are the main assumptions on which quantum
theory is founded.
In the canonical formalism these postulates 
are used together with the assumption that the time 
evolution of a state vector is governed by the
Schr{\"o}dinger equation.
In our study we will not use this assumption. 
Instead the following postulate 
will be  used:

(iii) The probability of an event is the absolute square of a 
complex number called the probability amplitude. 
The joint probability amplitude of a time-ordered 
sequence of events is product of the separate probability 
amplitudes of each of these events. 
The probability amplitude of an event which can happen in several 
different ways is a sum of the probability 
amplitudes for each of these ways.

The statements of the assumption (iii) express the well-known 
law for the quantum-mechanical probabilities. 
Within the canonical formalism this law is derived 
as one of the consequences of the theory. 
However, in the Feynman formulation of 
quantum theory this law is directly derived starting from 
the analysis of the phenomenon of quantum interference, 
and is used as a basic  postulate of the  theory.
We will also use the following assumption:

(iv) Under space-time translations $x \to x+a,$ the eigenstates $|n>$
of a system of non-interacting particles corresponding to the total
momentum of the particles ${\bf P}=\sum_i{{\bf p}_i}$ (${\bf p}_i$ is the
momentum of $i$- th particle) transforms as follows:
$$ |n> \tend_{x \tend x+a} exp(-iP_n a)|n>,$$
where $E_n= \sum_i{\sqrt{{\bf p}_i^2+m_i^2}}$ ($m_i$ are the
masses of the particles),
$P_n a \equiv P_n^{\mu}a_{\mu},$ and $a$ is an arbitrary 4-vector
of displacement, and $n$ stands for the entire set of discrete and
continuous variables that characterise the system in full.

Here and below, we use units where $\hbar =1.$ 
The statement of the assumption (iv) expresses the well-known 
law of transformation of vectors describing states of non-interacting 
particles under space-time translations. 
This law is used, for example, as one of the starting 
points in constructing the axiomatic approaches to 
quantum field theory (see, for example, Refs.[14,15]). 

The assumptions (i)-(iv) represent the current concepts of 
quantum theory. 
In the present paper we suggest a new way of using these concepts. 
In this way an equation of motion more general 
than the Schr{\"o}dinger equation will be derived as 
a consequence of the assumptions (i)-(iv) and the 
requirement of conservation of probabilities.  

\section*{III. Time-Evolution Operator}

Let $H_0$ be the free Hamiltonian, i.e. the operator of the total
energy of a system of particles travelling freely without
interaction or external disturbance. 
The vectors $|n>$ are the eigenvectors of this operator:
$H_0|n>= E_n|n>.$
As is well known, from the assumption (ii) it follows
that eigenvectors of any observable form a complete set
of basis vectors in the Hilbert space.
However, such continuum state vectors as the eigenvectors $|n>$
of the free Hamiltonian $H_0$ do not belong to the Hilbert
space of the states. 
Following the Dirac formalism we will
assume $|n>$ to be generalized basis vectors orthonormalized
in a continuum (by using delta functions) in terms of which
any vector of the Hilbert space can be expanded:
\begin{equation}
|\psi>=\sum_n |n><n|\psi>.
\end{equation}
From this and the
assumption (iv) it follows that if there is no interaction in
the system then any state vector has the
following time-dependence:
\begin{equation}
|\psi(t)> = exp(-iH_0t)|\psi(0)>.
\end{equation}
Below we will employ the interaction
representation in which the state vectors relate to the state vectors
in the Schr{\"o}dinger representation as follows:
$|\psi(t)>_I = exp(iH_0t)|\psi(t)>_S.$
From (2) it follows that in the interaction representation the state
vectors of non-interacting particles are independent of time.
Let us consider the probability amplitude 
$<\psi_2|U(t,t_0)|\psi_1>$
of finding, for a measurement at time $t,$ the quantum system in the
state $|\psi_2>$ if at time $t_0$ it was in the state $|\psi_1>.$
As it follows from the assumption (ii), this probability 
is the absolute square of the probability 
amplitude which is given by
the scalar product  $<\psi_2|\psi(t)>.$ 
Here $|\psi(t)>=U(t,t_0)|\psi_1>$ is the state at time $t$ to which
the system evolves from the state $|\psi_1>$ at  time $t_0$ 
if it is not disturbed by measurements; 
accordingly, $U(t,t_0)$ is
the operator describing such an evolution.
In the canonical approach to quantum theory $U(t,t_0)$ 
is postulated to be a unitary operator  
\begin{equation}
U^{+}(t_2,t_1) U(t_2,t_1) = U(t_2,t_1) U^{+}(t_2,t_1) = {\bf 1},
\end{equation}
with the group property
\begin{equation}
U(t_2,t_1) U(t_1,t_0) = U(t_2,t_0), \quad
U(t_0,t_0) = {\bf 1}.
\end{equation}

In the case of an isolated system (only such systems  will be  
considered in the present paper), the evolution operator in the  
Schr{\"o}dinger picture $U_s(t_2,t_1) \equiv exp(-iH_0t_2) U(t_2,t_1) 
exp(iH_0t_1)$ depends on the difference $(t_2-t_1)$ only, so that 
the operators $V(t) \equiv U_s(t,0)$  constitute a one-parameter group
of unitary operators, with the group property 
\begin{equation}
V(t_1+t_2) = V(t_1) V(t_2) , \quad    V(0)=0.
\end{equation}
If these operators are assumed to be strongly continuous, i.e. if 
\begin{equation}
\lim \limits_{t_2 \tend t_1} \Vert V(t_2)\vert\psi> - V(t_1) \vert \psi> 
 \Vert = 0,
\end{equation}
then from Stone's  theorem it follows [16] that this one-parameter 
group has a self-adjoint infinitesimal  generator $H$: 
$$ V(t) = exp(-iHt), \quad   
 i d/dt	V(t)=H V(t).$$
Identifying $H$ with the total Hamiltonian as usual, we get 
the time-dependent  Schr{\"o}dinger equation: 
$ i \frac{d |\psi_s(t)>}{dt} = H |\psi_s(t)>,$ 
where $ |\psi_s(t)> = V(t)|\psi_s(t=0)>.$
However, the condition (6) seems to be too strong.
From the physical point of view, it is 
enough to require that 
\begin{equation}
 <\psi_2| V(t_2)|\psi_1> \tend
 \limits_{t_2 \tend t_1}
 <\psi_2| V(t_1)|\psi_1> 
\end{equation}
for any physically realisable states $|\psi_1>$ and 
$|\psi_2>$  [14].
Note in this connection that there are normalized 
vectors in the Hilbert space that represent the 
states for which energy of a system is infinite. 
Such states cannot be considered as physically 
realisable [14], and hence the corresponding matrix 
elements of the evolution operator need not be continuous. 
For this reason, in the present paper we will not 
restrict ourselves to the strongly continuous evolution 
operators.
We will only assume that the evolution operator 
satisfies the condition (7). 
This means that we will not use the assumption that the time evolution 
of a state vector is governed by the Schr{\"o}dinger equation 
as the basic dynamical postulate. 

According to the assumption (iii), the probability amplitude of an 
event which can happen in several different ways is a 
sum of contributions from each alternative way. 
In particular, the amplitude 
 $<\psi_2| U(t,t_0)|\psi_1>$ can be represented as a sum 
of contributions from all alternative
ways of realisation of the corresponding evolution process.
Dividing these alternatives in different classes, we can then analyse
such a probability amplitude in different ways [13].
For example, subprocesses with definite instants of the
beginning and  end of the interaction in the system
can be considered as such alternatives.
Let $<n_2| U(t,t_0;t_2,t_1)|n_1>$ be the
probability amplitude that if at time $t_0$ the system was 
in the state $|n_1>,$ then the interaction in the system
will begin  at time $t_1$ and  end at time $t_2,$ and 
at time $t$ the system will be in the state $|n_2>.$
According to the postulate (iii), the amplitudes
$<n_2| U(t,t_0;t_2,t_1)|n_1>$
determine the contributions from the above alternatives, and
$<n_2|U(t,t_0)|n_1>$  can be represented as a sum
(more precisely as an integral) of these contributions
\begin{equation}
<n_2|U(t,t_0)|n_1>=  \varphi(n_2,n_1,t,t_0) + 
\int_{t_0}^t dt_2
\int_{t_0}^{t_2} dt_1 <n_2| U(t,t_0;t_2,t_1)|n_1>,
\end{equation}
where $\varphi (n_2,n_1,t,t_0)$ is the probability amplitude that 
if at time $t_0$ the system was in the state $|n_1>,$ then the 
particles of the system will  not interact in the
time interval $(t_0,t),$ and at time $t$ the 
system will be in the state $|n_2>$. 
Thus the first term on the right-hand side of (8) is the contribution 
from the alternative subprocess in the case of which the 
particles of the system do not interact. 
From the postulate (iii), it follows that the probability amplitude 
$<n_2| U(t,t_0;t_2,t_1)|n_1>$ is expressible as a product 
\begin{equation}
<n_2| U(t,t_0;t_2,t_1)|n_1>
= \sum_{n^{\prime}} \sum_n 
\varphi (n_2,n^{\prime},t,t_2) <n^{\prime}|\tilde S(t_2,t_1)|n>
\varphi (n,n_1,t_1,t_0),
\end{equation}
where $<n_2|\tilde S(t_2,t_1)|n_1>$ is the 
probability amplitude that if at time $t_1$ the 
system was in the state $|n_1>,$ then the 
interaction in the system will begin at time $t_1$ and 
will end at  time $t_2,$ and at this time the 
system will be in the state $|n_2>.$ 

The evolution operator $U(t,t_0)$ has a natural decomposition 
\begin{equation}
U(t,t_0) = {\bf 1} + i R(t,t_0).
\end{equation}
Here the unit operator  represents the no-interaction 
part; its matrix elements are delta functions which make 
the final momenta the same as the initial momenta. 
The operator $R(t,t_0)$ represents the interaction part. 
From (8) and (10) it follows that $\varphi (n_2,n_1,t_2,t_1) = <n_2|n_1>,$ 
and hence $<n_2| U(t,t_0;t_2,t_1)|n_1>= 
<n_2|\tilde S(t_2,t_1)|n_1>.$ 
Thus equation (8) can be rewritten in the form 
\begin{equation}
<n_2| U(t,t_0)|n_1> = <n_2|n_1> + 
\int_{t_0}^t dt_2 \int_{t_0}^{t_2} dt_1
<n_2|\tilde S(t_2,t_1)|n_1>;
\end{equation}
accordingly, the   
evolution operator $U(t,t_0)$ can be expressed 
in terms of the operators $\tilde S(t_2,t_1)$ whose matrix 
elements are the amplitudes $<n_2|\tilde S(t_2,t_1)|n_1>:$
\begin{equation}
 U(t,t_0) = {\bf 1} + 
\int_{t_0}^t dt_2 \int_{t_0}^{t_2} dt_1
\tilde S(t_2,t_1).
\end{equation}
Equation (12) defines the evolution operator $U(t_2,t_1)$ 
only for $t_2 \geq t_1.$ 
Using (3), the evolution operator 
$U(t_2,t_1)$  for $t_2 < t_1$ can be constructed as follows:
\begin{equation}
U(t_2,t_1)=
U^{-1}(t_1,t_2) = 
U^{+}(t_1,t_2).   
\end{equation}
It should be noted that in general $\tilde S(t_2,t_1)$ may be an 
operator-valued distribution, since only $U(t,t_0)$ has to be an 
operator on the Hilbert space. 
Nevertheless, throughout this paper we will use term "operator" 
for $\tilde S(t_2,t_1).$ 

To clarify the role which  the operator $\tilde S(t_2,t_1)$ 
plays in the proposed formalism, note the following. 
The  Feynman formulation is based on the assumption 
that the history of a system can be  represented by some path in 
space-time. 
From the postulate (iii) it then follows that the 
probability amplitudes of any event  is 
a sum of the probability amplitudes that a particle 
has a completely specified path in space-time. 
The contribution from a single path is postulated to be an 
exponential whose (imaginary) phase is the classical action 
(in units of $\hbar$) for the path in question. 
In the proposed formalism the history of a system is 
represented by the version  of the time evolution 
of the system associated with completely specified instants of 
the beginning and end of the interaction in the system.
Such a description of the history of a system is  
more general and requires no supplementary 
postulates like the above assumptions of the  Feynman formulation.
On the other hand, the probability  amplitudes 
$<\psi_2|\tilde S(t_2,t_1)|\psi_1>$ in terms of which we describe 
quantum dynamics are used in the spirit of Feynman's theory: 
The probability amplitude of any event is  
represented as a sum of these amplitudes. 
Bellow we will show that the requirement of unitarity 
for the evolution operator given by (12) leads 
to an equation for the operator $\tilde S(t_2,t_1)$  which 
can be regarded as an equation of motion. 
 
As we have noted, $<\psi(t_2)|\tilde S(t_2,t_1)|\psi(t_1)>$ 
is the probability amplitude that if at time 
$t_1$ the  system was in the state $|\psi_1>,$ 
then  the interaction in the system will begin 
at time $t_1$ and will end at time $t_2,$ and at 
this time the system will be found to be in the 
state $|\psi(t_2)>.$ 
Here this probability amplitude is represented by the 
matrix element $<\psi(t_2)|\tilde S(t_2,t_1)|\psi(t_1)>$ 
in the interaction picture. 
However, the same probability amplitude 
can be represented by the matrix element 
$<\psi_s(t_2)|\tilde S_s(t_2,t_1)|\psi_s(t_1)>$ 
in the Schr{\"o}dinger picture,
where $\tilde S_s(t_2,t_1)$ is the operator 
describing the transformation of a state at time 
$t_1$ into the state at time $t_2$ caused by 
the interaction in the system that begins at time 
$t_1$ and ends at time $t_2.$ 
Being an operator in the Schr{\"o}dinger picture,
$\tilde S_s(t_2,t_1)$  depends on the 
difference $(t_2 -t_1)$ only:
$\tilde S_s(t_2,t_1) \equiv \tilde T(t_2-t_1).$
Since $<\psi_s(t_2)|\tilde S_s(t_2,t_1)|\psi_s(t_1)>$ 
and $<\psi_s(t_2)|\tilde T(t_2-t_1)|\psi_s(t_1)>$ 
represent the same probability amplitude, we have
\begin{equation}
<\psi_s(t_2)|\tilde S_s(t_2,t_1)|\psi_s(t_1)> = 
<\psi_s(t_2)|\tilde T(t_2-t_1)|\psi_s(t_1)>.
\end{equation}
Taking into account the relation between the 
states in the Schr{\"o}dinger and interaction pictures
given by 
$|\psi(t)>=exp(iH_0t)|\psi_s(t)>,$ 
from (14), we get 
\begin{equation}
\tilde S(t_2,t_1) = exp(iH_0t_2)
\tilde T(t_2-t_1) exp(-iH_0 t_1)>.
\end{equation}

Let us now consider the scattering matrix. 
Letting $t_0 \to -\infty$ and $t \to \infty$ in (12), 
we get the S matrix in the following form:
\begin{equation}
<n_2|S|n_1> =<n_2|n_1> +
\int_{-\infty}^{\infty} dt_2 \int_{-\infty}^{t_2} dt_1
<n_2|\tilde S(t_2,t_1)|n_1>.
\end{equation}
By using (15), this expression can be rewritten 
in the form 
\begin{equation}
<n_2|S|n_1> = 
<n_2|n_1> - \quad  2\pi i \delta(E_{n_2}- E_{n_1}) <n_2| T(E_{n_1})|n_1>,
\end{equation}
where the operator $T(z)$ is defined by 
\begin{equation}
<n_2|T(z)|n_1> = i \int_{-\infty}^{\infty} d\tau exp(iz\tau)
 <n_2| F(\tau)|n_1>,
\end{equation}
with  
\begin{equation}
<n_2| F(\tau)|n_1>= <n_2|\tilde T(\tau)|n_1>, 
\quad  \tau \geq 0, 
\end{equation}
and $<n_2|F(\tau)|n_1> = 0 $ \quad  for \quad  $\tau<0.$
The fact that $F(\tau)=0$ for $\tau <0$ can be 
considered as an  expression of causality, since 
$\tau$ is assumed to be the duration time of interaction 
that has to be positive.
A remarkable consequence of this fact is 
that $<n_2|T(z)|n_1>$ is analytic in the 
upper half of the complex $z$ plane.

\section*{IV. Unitarity Condition and its Consequences }

Let us consider the unitarity condition that expresses the 
principle of conservation of probabilities. 
Substituting  (10) into the unitarity condition (3) gives 
$$
<n_2|R(t,t_0)|n_1> - <n_2|R^+(t,t_0)|n_1> =
$$
\begin{equation}
= i \sum_n <n_2|R^+(t,t_0)|n> <n|R(t,t_0)|n_1>,
\end{equation}
$$
<n_2|R(t,t_0)|n_1> - <n_2|R^+(t,t_0)|n_1> = 
$$
\begin{equation}
= i \sum_n <n_2|R(t,t_0)|n> <n|R^+(t,t_0)|n_1>.
\end{equation}
From (10),(12) and (15) it follows that the operator 
$R(t,t_0)$ can be written as   
\begin{equation}
R(t,t_0) =  
- i \int_{t_0}^{t} dt_2 \int_{t_0}^{t_2} dt_1
exp(iH_0 t_2) \tilde T(t_2-t_1)exp(-iH_0 t_1).
\end{equation}
Taking into account that, according to (18), 
the operator $\tilde T(\tau)$ 
can be represented in the form 
\begin{equation}
\tilde T(\tau) = - \frac{i}{2\pi} \int _{-\infty}^{\infty}dx
exp(-iz\tau) T(z),
\end{equation}
where $z=x+iy,$ $x$ and $y$ are real, and $y>0,$ we get 
$$<n_2|R(t,t_0)|n_1> = - \frac{1}{2\pi} \int _{t_0}^{t} dt_2
\int _{t_0}^{t_2}dt_1 \int^{\infty}_{-\infty} dx 
$$
\begin{equation}
\times exp(i E_{n_2} t_2) exp[-iz(t_2-t_1)] exp(-i E_{n_1} t_1)
<n_2|T(z)|n_1>. 
\end{equation}
Let us assume that $<n_2|T(z)|n_1>$ satisfies the condition 
\begin{equation} 
\lim \limits_{|z| \to \infty} 
\frac{<n_2|T(z)|n_1>}{z} = 0 .
\end{equation}
Note in this connection that the properties of $<n_2|T(z)|n_1>$ at 
infinity depend on the behaviour of $<n_2|\tilde S(t_2,t_1)|n_1>$ 
in the limit $t_2 \to t_1.$ 
As will be shown bellow, in the proposed formalism this 
behaviour determines the dynamics in the system. 
Thus, by assuming that $<n_2|T(z)|n_1>$ satisfies the condition 
(25), we restrict ourselves to considering some class of the 
dynamical schemes. 
However, the important thing for us is that, as it will be 
shown below, the Hamiltonian dynamics belongs to this class. 
Taking into account the condition (25), from (24), we get 
(see Appendix A)
$$
<n_2|R(t,t_0)|n_1> =  
$$
\begin{equation}
= \frac{1}{2\pi}  
\int^\infty_{-\infty} dx 
 \frac {exp[-i(z-E_{n_2})t] exp[i(z-E_{n_1})t_0]}
{(z-E_{n_2})(z-E_{n_1})}
<n_2|T(z)|n_1> .  
\end{equation}
Substituting of (26) into (20) gives 
$$
\int^\infty_{-\infty} dx_1  
 \frac {exp[-i(z_1-E_{n_2})t]exp[i(z_1-E_{n_1})t_0]}
{(z_1-E_{n_2})(z_1-E_{n_1})}
<n_2|T(z_1)|n_1> - 
$$
$$
- \int^\infty_{-\infty} dx_2  
 \frac {exp[i(z^*_2-E_{n_1})t]exp[-i(z_2^* -E_{n_2})t_0]}
{(z_2^* -E_{n_2})(z^*_2-E_{n_1})}
<n_2|T^+(z_2)|n_1> = 
$$
\begin{equation}
= <n_2|M(t,t_0)|n_1>,  
\end{equation}
where $z_1= x_1+iy,$ $z_2=x_2+ iy, y>0,$ and 
$$<n_2|M(t,t_0)|n_1> 
= \frac{i}{2\pi}  \int^\infty_{-\infty} dx_1
\int^\infty_{-\infty} dx_2  $$
\begin{equation}
\times \frac {exp[i(z^*_2-z_1)(t-t_0)]exp[i(E_{n_2}-E_{n_1})t_0]}
{(z^*_2-E_{n_2})(z_1-E_{n_1})}
<n_2|B(z_2,z_1)|n_1>,  
\end{equation}
with 
\begin{equation}
<n_2|B(z_2,z_1)|n_1> =    \sum_n 
\frac 
{<n_2|T^+(z_2)|n> <n|T(z_1)|n_1>}
{(z^*_2-E_n)(z_1-E_n)}. 
\end{equation}
Note that equation (27) derived in this way must be satisfied 
for $t \geq t_0.$
However, taking into account  the analytic properties of 
$<n_2|T(z)|n_1>,$ one can easily verify that 
both sides of (27) are equal to zero for $t<t_0.$ 
Thus equation (27) is valid for all $t$ and $t_0.$  
Represent the operator $B(z_1,z_2)$ in the form 
\begin{equation}
B(z_1,z_2) =  
\frac {T(z_1)- T^+(z_2)} {z_2^* - z_1} + 
F(z_1,z_2). 
\end{equation}
Substituting (30) into (27) and taking into account the analytic 
properties of $<n_2|T(z)|n_1>,$ we have then 
 $$
 \int^\infty_{-\infty} dx_1
\int^\infty_{-\infty} dx_2  
 \frac {exp[i(z^*_2-z_1)(t-t_0)]exp[i(E_{n_2}-E_{n_1})t_0]}
{(z^*_2-E_{n_2})(z_1-E_{n_1})}
$$
\begin{equation}
\times <n_2|F(z_1,z_2)|n_1> = 0.   
\end{equation}
Since equation (31) must be satisfied for all $t$ and $t_0,$  we get 
\begin{equation}
<n_2|F(z_1,z_2)|n_1> = 0.   
\end{equation}
From this it follows that 
\begin{equation}
T(z_1) - T^+(z_2) = (z^*_2 -z_1) \sum_n 
 \frac {T^+(z_2)|n><n|T(z_1)}
{(z^*_2-E_n)(z_1-E_n)} .   
\end{equation}
Proceeding in an analogous way, from (21), we obtain 
\begin{equation}
T(z_1) - T^+(z_2) = (z^*_2 -z_1) \sum_n 
 \frac {T(z_1)|n><n|T^+(z_2)}
{(z^*_2-E_n)(z_1-E_n)} .   
\end{equation}
It is easy to show then that the following equation 
results from (33) and (34):  
\begin{equation}
T(z_1) - T(z_2) = (z_2 -z_1) \sum_n 
 \frac {T(z_2)|n><n|T(z_1)}
{(z_2-E_n)(z_1-E_n)} .   
\end{equation}
Thus we have shown that for the evolution operator 
given by (12)  to be unitary the operator $T(z)$ 
must satisfy equations (33), (34) and (35).
At the same time, as it shown in Appendix B, the evolution
operator given by (12) satisfies the composition law (4),
provided equation (35) is valid.
Note that equation (35) can be derived from the composition law 
(4) and the representation (12) in the same way in which 
(33) has been derived from this representation and the unitarity 
condition (20). 
However, equations (33) and (34) are more general than equation (35).
In fact, (35) results from (33) and (34). 
But for (33) and (34) to result from (35) the additional 
assumption that $T^+(z)=T(z^*)$ is required.

Equation (35) plays a key  role in the proposed formalism.
It has been derived as a consequence of such 
fundamental physical principles of quantum theory as the 
requirement of conservation of probabilities and 
the principle of the superposition of probability amplitudes. 
It should be noted that an equation having the same form as 
equation (35) has been derived 
for the scattering T matrix 
as a consequence of the above 
principles within the approach to the scattering theory 
developed in Refs.[17-20]. 
That equation was derived as an off-shell generalization of the 
unitarity condition for the scattering S matrix, and may be 
regarded as a particular case of equation (35). 

\section*{V. Equation of Motion for State Vectors}

In the previous section we have derived equations (33), (34) and (35) 
for the operator $T(z).$ 
It now becomes necessary for us to obtain an equation 
directly for the operator $\tilde S(t_2,t_1)$ 
which is of central importance in our formalism.
It can be shown (see Appendix C) that for 
equation (35) to be valid for any $z_1$ and $z_2$ 
the operator must satisfy the following equations:
$$
(1- exp[ia(t_2-t_1)]) 
\tilde S(t_2,t_1) = 
$$
\begin{equation}
= \int^{t_2}_{t_1} dt_4
\int^{t_4}_{t_1}dt_3 exp[ia(t_2-t_4)] 
(1- exp[ia(t_4-t_3)])
\tilde S(t_2,t_4) \tilde S(t_3,t_1),
\end{equation}
\begin{equation}
(t_2-t_1) \tilde S(t_2,t_1) = 
\int^{t_2}_{t_1} dt_4 \int^{t_4}_{t_1}dt_3
(t_4-t_3) \tilde S(t_2,t_4) \tilde S(t_3,t_1),
\end{equation}
where $a=z_2-z_1.$
Substituting (15) into (37), we obtain then the corresponding 
equation for the operator $\tilde T(\tau)$
\begin{equation}
\tau \tilde T(\tau) = 
\int^\tau_0 d\tau_2  \int_0^{\tau-\tau_2} d\tau_1 (\tau-\tau_1-\tau_2)
\tilde T(\tau_2)exp[-i(\tau-\tau_1-\tau_2)H_0]
\tilde T(\tau_1).
\end{equation}
It should be noted that (38) coincides in form with the equation 
for the operator $\tilde T(\tau)$ derived within scattering theory [21]. 
However, (38) was initially considered as an equation 
allowing one to find out what boundary conditions for 
equation (35) are admissible. 
In Ref.[22]  it was shown 
that this equation enables one to construct a new class models in 
nonrelativistic quantum scattering theory, which were shown to open 
new possibilities for describing hadron-hadron interactions 
at low energies. 

As we will show below, equation (37) allows one to obtain  
the amplitudes $<n_2|\tilde S(t_2,t_1)|n_1>$
for any $t_1$ and $t_2$ if the amplitudes
$<n_2|\tilde S(t'_2, t'_1)|n_1>$ corresponding to infinitesimal 
duration times $\tau = t'_2 -t'_1$ of interaction are known.
It is natural to assume that most of the 
contribution to the evolution operator 
in the limit $t_2 \to t_1$ comes from the processes 
associated with an fundamental interaction  
in the system under study.
Denoting this contribution by 
$H_{int}(t_2,t_1)$, the operator 
$\tilde S(t_2,t_1)$ can be represented in the form 
\begin{equation}
\tilde{S}(t_2,t_1) = 
H_{int}(t_2,t_1) + \tilde{S}_1(t_2,t_1), 
\end{equation}
where $\tilde{S}_1(t_2,t_1)$ is the part of the 
operator $\tilde{S}(t_2,t_1)$ which in the limit 
$t_2 \tend t_1$ gives the 
negligibly small contribution to the evolution 
operator  little in comparison with $H_{int}(t_2,t_1).$
We will assume that the 
operator $H_{int}(t_2,t_1)$ contain all the 
dynamical information that is needed to 
construct the evolution operator $U(t_2,t_1).$ 
Thus we will assume the operator $H_{int}(t_2,t_1)$ 
to play the role which the interaction 
Hamiltonian plays in the ordinary formulation of quantum theory:
It generates dynamics in a system. 
This operator can be regarded as a generalization of the 
interaction Hamiltonian, and we will call it  the
generalized interaction operator.
Obviously, the operator  $H_{int}(t_2,t_1)$
must satisfy (37) in the limit $t_2 \to t_1$ 
\begin{equation}
F_1(t_2,t_1) \tend \limits_{t_2 \to t_1} 0,
\end{equation}
where
$$
F_1(t_2,t_1) = 
- (t_2-t_1) H_{int}(t_2,t_1) +
\int^{t_2}_{t_1} dt_4 \int^{t_4}_{t_1} dt_3 
(t_4-t_3) H_{int}(t_2,t_4) H_{int}(t_3,t_1).
$$
According to (15), the operator $H_{int}(t_2,t_1)$ can be 
represented in the form
\begin{equation}
H_{int}(t_2,t_1) = exp(iH_0t_2) H_{int}^{(s)}(t_2-t_1)
exp(-iH_0t_1),
\end{equation}
where $H^{(s)}_{int}(t_2-t_1)$ is the generalized 
interaction operator in the Schr{\"o}dinger picture. 
If  $H_{int}(t_2,t_1)$ is specified, equation (37) allows one to find the
operator $\tilde S(t_2,t_1).$ 
Formula (12) can then be used to construct the evolution operator 
$U(t,t_0)$ and accordingly the state vector 
\begin{equation}
|\psi(t)> = |\psi(t_0)> +  \int_{t_0}^t dt_2
\int_{t_0}^{t_2} dt_1 \tilde S(t_2,t_1) |\psi(t_0)> 
\end{equation}
at any time $t.$ 
Thus (37) can be regarded as an equation of motion for states 
of a quantum system.

From the  mathematical point of view the 
requirement that $H_{int}(t_2,t_1)$ contains all the 
dynamic information that is needed for constructing 
$U(t_2,t_1)$ means that the operator $H_{int}(t_2,t_1)$ 
must have such a form that the equation (37) 
has a unique solution having the following behaviour 
near the point $t_2=t_1:$
\begin{equation}
\tilde S(t_2,t_1)
\tend \limits_{t_2 \tend t_1}
 H_{int}(t_2,t_1) 
+ o(\tau^{\epsilon}),
\end{equation}
where $\tau=t_2-t_1$ and the value of $\epsilon$ 
depends on the form of the operator $ H_{int}(t_2,t_1).$
In order to clarify this point,  
note that equation (38) is equivalent to 
the following equation for the 
operator  $T(z)$ given by (18): 
\begin{equation}
\frac{d T(z)}{dz} =
 T(z) G^{(2)}(z) T(z) ,
\end{equation}
where 
$$
 G^{(2)}(z) = - \sum \limits_{n} 
\frac{|n><n|}{(z-E_n)^2} .
$$

The correspondence between equation (38) and the 
differential equation (44) can be easily stated 
in the same way in which we have stated correspondence 
between equations (35) and (36). 
Thus, instead of solving (37) or (38), one can solve 
equation (44) for the operator $T(z).$ 
The operator $\tilde T(\tau)$ and correspondingly 
the operator $\tilde S(t_2,t_1)$ can then be 
obtained by using (23). 
At the same time, according to (26), the operator 
$T(z)$ can be dirrectly used for constructing 
the evolution operator. 
According to (18), (39) and (41), the operator $T(z)$ 
has the following asymptotic behaviour 
for $|z| \tend \infty:$ 
\begin{equation}
T(z) \tend \limits_{|z| \tend \infty}
 B(z) + o(|z|^{-\beta}) ,
\end{equation}
where
\begin{equation}
B(z) = i \int_0^{\infty} d\tau exp(iz \tau)
H^{(s)}_{int}(\tau),
\end{equation}
and $\beta=1+\epsilon.$  
From (44) and (45) it follows that the operator $B(z)$ must 
satisfy the following asymptotic condition:
\begin{equation}
\frac{d B(z)}{dz} 
\tend \limits_{|z| \tend \infty}
 B(z) G^{(2)}(z) B(z)+ o(|z|^{-\beta}).
\end{equation}
The above requirements, which the operator 
$H^{(s)}_{int}(\tau)$ has to meet, mean that 
$B(z)$ must be so close to the solution of 
equation (44) in the limit $|z| \tend \infty$ that this 
differential equation has a unique solution 
having the asymptotic behaviour (45). 
The operator $B(z)$ represents the contribution 
which $H^{(s)}_{int}(\tau)$ gives to the 
operator $T(z),$ and we will call it the 
effective interaction operator.

Let us now show that the 
Schr{\"o}dinger equation results from equation (37) 
and hence the dynamics of a quantum system is 
equivalent to the Hamiltonian dynamics in the 
case when the generalized interaction 
operator is of the form 
\begin{equation}
 H^{(s)}_{int}(\tau) = - 2i \delta(\tau) 
 H_{I} ,
\end{equation}
$H_{I}$ being a self-adjoint operator. 
The delta function $\delta(\tau)$ in (48) 
emphasises the fact that in this case, 
the fundamental interaction is instantaneous. 
As it is shown in Appendix C, from (37) it 
follows that 
$\tilde S(t_2,t_1)$ must satisfy also equation (36) for any $a.$
Letting $a \to i \infty$ in (36) and taking 
into account (48), 
we get the following equation:
$$
<n_2| \tilde{S}(t_2,t_1) |n_1> = 
- 2 i \delta (t_2-t_1) <n_2| H_I(t_1)|n_1> 
$$
\begin{equation}
- i \sum \limits _{n}   
\int^{t_2}_{t_1} dt_3 <n_2| H_I(t_2) |n><n| 
\tilde S(t_3,t_4) |n_1>, 
\end{equation}
where $H_{I}(t) = exp(iH_0 t) H_{I} exp(-i H_0 t).$ 
Solving this equation by expanding 
$\tilde S(t_2,t_1)$ in terms of $H_I(t),$ 
one can easily get
$$
\tilde S(t,t_0)
 = - 2i \delta(t-t_0) 
 H_{I}(t_0) +
$$
\begin{equation}
+ \sum_{n=2}^{\infty}(-i)^n
\int^{t}_{t_0} dt_1 \int^{t_1}_{t_0}dt_2
\cdots 
\int^{t_{n-3}}_{t_0} dt_{n-2}
 H_{I}(t)  H_{I}(t_1) \cdots 
 H_{I}(t_{n-2}) H_{I}(t_0).
\end{equation}
Thus we have shown that for the case under study 
equation (37) allows one to obtain $\tilde S(t_2,t_1)$ for 
any $t_1$ and $t_2$ starting from the contribution 
to the evolution operator coming from the process 
associated with the instantaneous interaction. 

Inserting (50) into (12) yields 
\begin{equation}
U(t,t_0) = {\bf 1} 
+ \sum_{n=1}^{\infty}(-i)^n
\int^{t}_{t_0} dt_1 
\cdots 
\int^{t_{n-1}}_{t_0} dt_{n}
 H_{I}(t_1)  H_{I}(t_2) \cdots 
 H_{I}(t_{n}).
\end{equation}
This expression  coincides in form with the 
Dyson expansion of the evolution operator. 
From this it follows that the operator $H_I(t)$ 
has to be identified with the interaction Hamiltonian. 
As is well known, the series (51) is convergent, 
provided the operator $H_I(t)$ is bounded.
This means that the series (50) is also convergent in this case. 
On the other hand, substituting (49) into (12), we easily get 
the following equation for the evolution operator 
$U(t_2,t_1):$
\begin{equation}
U(t_2,t_1) = {\bf 1} 
- i \int^{t_2}_{t_1} dt 
 H_{I}(t) U(t,t_1) .
\end{equation}
This equation is the integral form of the 
Schr{\"o}dinger equation for the evolution operator
\begin{equation}
 \frac {dU(t,t_0)}{dt} = -i
 H_{I}(t) U(t,t_0) , \quad
U(t_0,t_0)= {\bf 1} .
\end{equation}
Hence in the case when 
$H_{int}^{(s)}(\tau)$ is of the 
form (48), from (37) it follows that $|\psi(t)>$ 
given by (42) satisfies the 
Schr{\"o}dinger equation
$$
  \frac {d |\psi(t)>}{d t} = -iH_I(t)|\psi(t)>.
$$
Thus we have shown the equivalence of the 
dynamics governed by equation (37) to the Hamiltonian 
dynamics in the case when the generalized 
interaction operator $H_{int}^{(s)}(\tau)$ 
is of the form (48). 
At the same time, as will be shown below, equation (37) 
permits the generalization to the case when the operator
$H_{int}^{(s)}(\tau)$ has no such a singularity at the point 
$t_2=t_1$ as the delta function.
In this case the fundamental interaction 
generating the dinamics in a quantum system is 
nonlocal in time:
The evolution operator is defined by 
$H_{int}^{(s)}(\tau)$ as a function of the time 
duration $\tau$ of the interaction.

It should be noted that the concept of nonlocal 
in time potentials was first introduced within the 
optical-potential model. 
This concept, for example, is used in the theory 
of time-dependent optical potentials (see, for 
example, [23,24] and references therein). 
The optical potentials are introduced in the case 
when only state vectors belonging some subspace 
of the Hilbert space are included explicitly in the 
description of the  time evolution of a quantum system. 
Such potentials which account for the coupling 
between this subspace and its complementary part of 
the Hilbert space are nonlocal in time and, 
thus, depends on the history of a dynamical system. 
The nonlocal form of the optical potentials 
is an expression of the loss of probability from the 
above subspace.
However, the optical-potential model is one 
of the methods for the description of the time 
evolution of quantum systems the dynamics of which is 
generated by the local in time interaction being 
described by the interaction Hamiltonian. 
In the present formalism such a dynamics corresponds 
to the particular case when the generalized interaction 
operator is of the form (48). 
At the same time our formalism permits the generalization 
to the case when the fundamental interaction generating 
the dynamics in a quantum system itself is nonlocal 
in time and hence the nonlocality of the 
interaction does not lead to the loss of probability 
in the system. 

\section*{VI. Exactly Solvable Model}

Let us consider the evolution problem for two nonrelativistic particles 
in the c.m.s. 
We denote the relative momentum by ${\bf p}$ and 
the reduced mass by $\mu.$ 
Assume that the generalized interaction operator in the 
Schr{\"o}dinger picture $H^{(s)}_{int}(\tau)$ 
has the form 
\begin{equation}
<{\bf p}_2| H^{(s)}_{int}(\tau) |{\bf p}_1> = 
\varphi({\bf p}_2) \varphi^{*}({\bf p}_1) f(\tau),
\end{equation} 
where $f(\tau)$ is some function of $\tau,$ and 
the form factor $\varphi ({\bf p})$ has the 
following asymptotic behaviour for $|{\bf p}| \tend \infty:$
\begin{equation}
\varphi({\bf p}) \sim \frac {c_1} {|{\bf p}|^{\alpha}}, 
\quad  {(|{\bf p}| \tend \infty).} 
\end{equation}
In this case, the problem can be easily solved by 
using equation (44). 
Representing $<{\bf p}_2| T(z) |{\bf p}_1>$ in the form 
\begin{equation}
<{\bf p}_2| T(z)|{\bf p}_1> = 
\varphi ({\bf p}_2)\varphi^* ({\bf p}_1) t(z),
\end{equation}
from  (44) and (45), we get the equation 
\begin{equation}
\frac {dt(z)}{dz} = 
-t^2(z) \int d^3k \frac {|\varphi ({\bf k})|^2}
{(z-E_k)^2}
\end{equation}
with the asymptotic condition 
\begin{equation}
t(z)  \tend \limits_{|z| \tend -\infty} 
f_1(z) + o(|z|^{-\beta}),
\end{equation}
where 
\begin{equation}
f_1(z)= i \int_0^{\infty} d\tau 
exp(iz\tau) f(\tau),
\end{equation}
and $E_k = \frac {k^2}{2 \mu}.$ 
The solution of equation (57) 
with the initial condition $t(a)=g_a,$ 
where $a \in (-\infty,0),$ is 
\begin{equation}
t(z) = g_a \left(1 +(z-a) g_a 
 \int d^3k \frac {|\varphi ({\bf k})|^2}
{(z-E_k)(a-E_k)} \right)^{-1}.
\end{equation}
In the case $\alpha > 1/2,$ the 
function $t(z)$ tends to a constant as 
$z \tend -\infty$
\begin{equation}
t(z)  \tend \limits_{z \tend -\infty}  
\lambda ,
\end{equation}
where 
\begin{equation}
\lambda = g_a \left(1 + g_a 
 \int d^3k \frac {|\varphi ({\bf k})|^2}
{a-E_k} \right)^{-1}.
\end{equation}
Thus in this case the function $f_1(z)$ must 
tend to $\lambda$ as $z \tend -\infty.$ 
From this and (59) it follows that the only 
possible form of the function $f(\tau)$ is 
\begin{equation}
f(\tau) = -2i \lambda \delta(\tau) 
+ f^{\prime}(\tau),
\end{equation}
where the function $f^{\prime}(\tau)$ 
has no such a singularity at the point $\tau=0$ 
as the delta function.
In this case  the generalized interaction operator 
$H^{(s)}_{int}(\tau)$ has the form (48) and hence the 
dynamics generated by this operator is equivalent 
to the dynamics governed by the 
Schr{\"o}dinger equation with the separable potential
\begin{equation}
<{\bf p}_2|H_I|{\bf p}_1> = \lambda \varphi({\bf p}_2) 
\varphi^*({\bf p}_1).
\end{equation}
In particular, in this case from (60) and 
(61) we easily get the well-known expression for the 
T matrix in the separable-potential model
\begin{equation}
<{\bf p}_2|T(z)|{\bf p}_1> = 
\lambda \varphi ({\bf p}_2)\varphi^{*}({\bf p}_1) 
\left (1 - \lambda \int d^3k \frac 
{|\varphi({\bf k})|^2}{z-E_k} \right )^{-1}. 
\end{equation}

Ordinary quantum mechanics does not permit the 
extension of the above model to the case 
$\alpha \leq1/2.$
Indeed, in the case of such a large-momentum behaviour of 
the form factors $\varphi({\bf p}),$ 
substituting the interaction Hamiltonian given by (64) 
into (51) leads to the ultraviolet divergences, and 
the integral in (65) is not convergent. 
We will now show that our formalism allows one to 
extend this model to the case $0 < \alpha <1/2.$
Let us determine the class of the functions $f_1(z)$ 
and correspondingly the value of $\beta$ for 
which equation (57) has a unique solution having the 
asymptotic behaviour (58).
In the case $\alpha <1/2,$ the 
function $t(z)$ given by (60) has the following 
behaviour for $z \tend -\infty:$
\begin{equation}
t(z)  \tend \limits_{z \tend -\infty}  b_1 |z|^{\alpha-1/2}+
b_2 |z|^{2 \alpha-1} + o(|z|^{2 \alpha-1}),
\end{equation}
where 
$b_1 =- 1/2 cos(\alpha \pi) \pi^{-2} c_1^{-2} 
(2 \mu)^{\alpha-3/2}$ and $b_2= b_1 |a|^{1/2-\alpha} -
b_1^2 g_a^{-1}.$
Here and below we restrict ourselves to the case when 
$\varphi ({\bf p})= c_1 |{\bf p}|^{-\alpha}.$ 
The parameter $b_1$ does not depend on $g_a.$ 
This means that all solutions of equation (57) have the same 
leading term  in (66), and only the second term distinguishes 
the different solutions of this equation. 
Thus in order to obtain a unique solution of equation (57) 
we must specify the first two terms in the 
asymptotic behaviour of $t(z)$ for $z \tend - \infty.$ 
From this it follows that the functions 
$f_1(z)$ must be of the form 
\begin{equation}
f_1(z) = b_1 |z|^{\alpha-1/2} +
b_2 |z|^{2 \alpha -1} ,
\end{equation}
and $\beta=2 \alpha -1.$ 
Correspondingly the functions $f(\tau)$ must be of the form
\begin{equation}
f(\tau) = a_1 \tau^{-\alpha-1/2} +
a_2 \tau^{-2 \alpha} ,
\end{equation}
with $a_1= -ib_1 \Gamma ^{-1}(1-2\alpha) exp[i(-\alpha/2+
1/4) \pi],$ and 
$a_2= -b_2 \Gamma ^{-1}(1-2\alpha) exp(-i \alpha \pi),$
where $\Gamma(z)$ is the gamma-function.
This means that in the case 
$\alpha <1/2,$ the generalized interaction operator 
must be of the form 
\begin{equation}
<{\bf p}_2| {H}^{(s)}_{int}(\tau)|{\bf p}_1> = 
a_1 \varphi ({\bf p}_2)\varphi^* ({\bf p}_1) \tau^{-\alpha-1/2} +
a_2 \varphi ({\bf p}_2)\varphi^* ({\bf p}_1) \tau^{-2 \alpha}.
\end{equation}
Using (56) and (60), for $<{\bf p}_2| T(z)|{\bf p}_1>$,  we get
\begin{equation}
<{\bf p}_2| T(z)|{\bf p}_1> = 
N(z) \varphi ({\bf p}_2)\varphi^* ({\bf p}_1), 
\end{equation}
with
\begin{equation}
N(z) = g_a \left (1 + (z-a) g_a \int d^3 k
\frac{|\varphi({\bf k})|^2}
{(z-E_k)(a- E_k)} \right )^{-1},
\end{equation}
where  
$$g_a = \frac {b_1^2} 
{b_1 |a|^{1/2 -\alpha} + a_2 \Gamma(1-2\alpha)
exp(- i \alpha \pi)}.$$ 
It can be easily checked that 
$N(z)$  given by (71) does not depend on the 
choice of  the parameter $a,$ and we can let $a \tend -\infty.$ 
Taking into account that 
$$g_a\tend\limits_{a\to -\infty} b_1 |a|^{\alpha-\frac{1}{2}}+
b_2 |a|^{2\alpha-1}+o(|a|^{2\alpha-1}), $$ 
and letting $a\to -\infty$ in (71), we get
\begin{equation}
N(z)=\frac {b_1^2}{ b_1(-z)^{1/2-\alpha}-b_2}.
\end{equation}
In order that the operator $T(z)$ given by (70) 
and (72) satisfy equations (33) and (34) the 
parameter $b_2$ must be real. 
The same formula for $T(z)$ can be obtained by using equation (37).
It can be shown that if $H_{int}(t_2,t_1)$ is of the 
form (69), then equation (37) has a unique solution 
having the behaviour (43) with $\epsilon =-2 \alpha.$  
In the case when $\varphi ({\bf p})= c_1 |{\bf p}|^{-\alpha},$ 
this solution written for $\tilde T(\tau)$ is 
$$
<{\bf p}_2|\tilde T(\tau)|{\bf p}_1> = 
\sum \limits_{n=1}^{\infty} a_n \tau^{n(1/2-\alpha)-1}
\varphi({\bf p}_2)\varphi^{*}({\bf p}_1),
$$
where 
$$
a_n = a_2^{n-1}a_1^{2-n} \Gamma^{n-1}(1-2\alpha)
\Gamma^{2-n}(1/2-\alpha) 
\Gamma^{-1}(n/2-n \alpha). 
$$
Substituting this series into (18) we get 
$<{\bf p}_2| T(z)|{\bf p}_1>$ which can be 
represented in the form (70).
By using (10), (26) and (70), we can construct the 
evolution operator 
$$
<{\bf p}_2|U(t,t_0)|{\bf p}_1> = <{\bf p}_2|{\bf p}_1>
 - \frac {i}{2\pi} 
\int_{-\infty}^{\infty} dx
$$
\begin{equation}
\times 
\frac {exp[-i(z-E_{p_2})t] exp[i (z - E_{p_1})t_0]}
{(z-E_{p_2})(z-E_{p_1})} 
N(z) \varphi({\bf p}_2) \varphi^*({\bf p}_1), 
\end{equation}
where  $z=x+iy,$ and $y>0.$ 
Since $<{\bf p}_2|T(z)|{\bf p}_1>$ 
given by (70) satisfies equations (33), (34) and (35), the 
evolution operator $U(t,t_0)$ defined by (12) is a unitary 
operator satisfying the composition law (4).

We have stated the correspondence between the form 
of the generalized interaction operator and 
the large-momentum behaviour of the form 
factor $\varphi({\bf p}).$ 
In the case $\alpha >1/2,$ the operator 
$H^{(s)}_{int}(\tau)$ would necessarily have 
the form (48). 
In this case the fundamental interaction is 
instantaneous. 
In the case $0 <\alpha <1/2$ (the restriction $\alpha >0$ 
is necessary for the integral in (60) to be convergent), 
the only possible form of 
$H^{(s)}_{int}(\tau)$ is (69), and hence the interaction 
generating  the dynamics of the system is 
nonlocal in time.
In the Appendix D it is shown that in the case  
there are normolized vectors in the Hilbert space for which 
$<\psi_2|U(t,0)|\psi_1>$ are not continuous at $t=0.$
These vectors represent the states with infinite energy 
which are not physically realisable and hence the 
condition (7) is not violated. 
Nevertheless,  the evolution operator 
$V(t)= U_s(t,0)$ is not continuous and hence the 
group of these operators has no infinitesimal 
generator in this case. 
From this it follows that in this case the time evolution 
of a state vector is not governed by the 
Schr{\"o}dinger equation.
The cause of this discontinuity is quite obvious. 
Indeed, from the point of view of the states with 
infinite energy any time interval $\delta t$ 
is infinite and hence the corresponding matrix elements 
of the evolution operator $U(\delta t,0)$ 
must be independent of $\delta t,$ i.e. 
must be constant. 
As it is shown in Appendix D, in the case $\alpha >1/2$ 
such matrix elements of $U(t,0)$ 
being independent of $t$ are zero, and the discontinuity 
problem does not appear. 
This problem appears in the case $\alpha <1/2,$ 
i.e. in the case when the matrix elements  
$<{\bf k}_2|U(t_2,t_1)|{\bf k}_1>$ as functions of ${\bf k}_1$ 
and  ${\bf k}_2$ do not meet the requirements of 
ordinary quantum mechanics. 
Thus the cause of the above mentioned lack of 
continuity of the evolution operator is the "bad" 
large-momentum behaviour of the form factors $\varphi ({\bf p}).$ 
 
Let us now show that the evolution operator defined by (71) and (73) 
satisfies the condition (7). 
Using (71) and (73), for the matrix element of the evolution 
operator in the Schr{\"o}dinger picture 
$ V(t),$ we can write 
\begin{equation}
<\psi_2| V(t)|\psi_1> = <\psi_2|\psi_1> 
 - \frac {i}{2\pi} \int_{-\infty}^{\infty} dx
exp(-izt) \frac {b^2_1 J(z)}{b_1(-z)^{1/2-\alpha} -b_2},
\end{equation}
with 
\begin{equation}
J(z) =  \int d^3k_1 \int d^3k_2 
\frac {|c_1|^2 \psi_1({\bf k}_1)\psi^*_2({\bf k}_2)}
{|{\bf k}_1|^{\alpha} |{\bf k}_2|^{\alpha} 
(z-E_{k_1})(z-E_{k_2})},
\end{equation}
where  $z=x+iy,$  $y>0,$ and $\psi_i({\bf k})= <{\bf k}|\psi_i>,$ 
$i=1,2.$ 
By using the fact that the functions $\psi_1({\bf k})$ and 
$\psi_2({\bf k})$ must be square-integrable, it can be 
shown that, if the vectors $|\psi_1>$ and $|\psi_2>$ do 
not represent the above states with infinite energy, then 
the function $J(z)$ given by (75) is estimated as 
$$
J(z) \leq C(1 + |z|)^{-1/2-\alpha-\epsilon},
$$
where $\epsilon >0.$
From this and the properties of Fourier integrals it 
follows that for the physically realisable states the matrix 
elements  
$<\psi_2| V(t)|\psi_1>$ given by (74) are continuous 
functions of time for $-\infty<t<\infty$ and hence the evolution operator 
devined by (71) and (73) satisfies the condition (7). 

\section*{VII. Summary and Discussions}

We have constructed a formalism based on the above assumptions of the 
Feynman formulation and canonical quantum theory. 
By using the assumption (iii) in the spirit of the 
Feynman formalism, within the formalism the probability 
amplitude 
$<\psi_2| U(t,t_0)|\psi_1>$ is represented as a sum of 
contributions from the evolution processes associated
with completely specified instants of the beginning 
and end of the interaction in a quantum system, 
and $<\psi_2| \tilde S(t_2,t_1)|\psi_1>$
represents the contribution from the process 
in the case of which the interaction begins at time 
$t_1$ and ends at time $t_2.$
We have shown that this representation  
and the requirement  of unitarity for the 
evolution operator  leads to equation (37) for 
the operator $\tilde S(t_2,t_1)$ and equations (33),(34) and (35) 
for the operator $T(z)$ defined by (18). 
Equation (37)  allows 
one to obtain the operator 
$\tilde S(t_2,t_1)$ if the generalized 
interaction operator ${\cal H}_{int}(t_2,t_1)$ which 
determines the behaviour of $\tilde S(t_2,t_1)$ 
in the limit $t_2 \to t_1$ is specified.
It has been shown that the evolution operator constructed 
in this way is a unitary operator with the group 
property (4). 
In the case when $H^{(s)}_{int}(\tau)$ is of the 
form (48), the Schr{\"o}dinger equation 
has been shown to result from equation (37) 
which can be regarded as an equation of motion. 
In this case the fundamental interaction in the 
system is instantaneous and 
the dynamics governed  by this equation of motion 
is equivalent to the Hamiltonian dynamics. 
At the same time, our formalism permits the 
generalization to the case when the fundamental 
interaction in a quantum system being described by the 
operator  $H^{(s)}_{int}(\tau)$ is nonlocal in time. 
In this case the dynamics is not equivalent to the 
dynamics governed by the Schr{\"o}dinger equation. 
This is not at variance with Stone's theorem, 
since, as it has been shown on the exactly solvable 
model, the evolution operator is not strongly 
continuous in this case. 
It satisfies only the more general continuity 
condition (7). 
We have stated correspondence between the large-momentum 
behaviour of the matrix elements of the evolution 
operator and the form of the generalized interaction 
operator $H^{(s)}_{int}(\tau):$ 
If this behaviour satisfies the requirements of conventional 
quantum theory, then $H_{int}^{(s)}(\tau)$ must necessarily 
be of the form (48), and if this large-momentum 
behaviour does not 
meet the above requirements, then $H^{(s)}_{int}(\tau)$ 
must be of the form corresponding to the case 
when the interaction generating the dynamics 
of a quantum system is nonlocal in time. 

The above gives reason to hope that the proposed 
formalism may open new possibilities for solving 
the problem of the ultraviolet divergences in 
quantum field theory. 
In this connection note the following. 
Since locality is  the main cause of the 
ultraviolet divergences, it seems natural to 
resolve the problem by the introduction of a nonlocal 
form factor into the interaction Hamiltonian density. 
For example, in QED a nonlocal form factor could be 
introduced by specifying the interaction Hamiltonian 
density in the form 
$$
{\cal H}_I(x) = \int d^4 x_1 \int d^4 x_2
F(x-x_1, x-x_2) j_{\mu}(x_1) A^{\mu}(x_2)
$$
where
$F(x-x_1, x-x_2)$ is a covariant form factor,
$j_{\mu}(x)$ is the current density operator, and $A_{\mu}(x)$
is the electromagnetic field potential.
However, it turned out that the nonlocal form factor cannot
be introduced in such a way without losing  covariance
of the theory.
Indeed, for the theory to be relativistically invariant, 
${\cal H}_1(x)$ must satisfy the condition 
$$
{\cal H}_1(x_1){\cal H}_1(x_2)= {\cal H}_I(x_2) {\cal H}_I(x_1),
\quad  \quad      (x_2 - x_1)^2 <0. $$
But, as is well known, the introduction of the nonlocal form 
factor leads to violating this condition. 
The reason for these difficulties is quite obvious. 
Due to the delta function $\delta(\tau)$ the generalized 
interaction operator given by (48) describes an instantaneous 
interaction. 
In nonrelativistic quantum mechanics processes of 
instantaneous interaction may be nonlocal in space. 
However, in relativistic quantum theory a local in time 
process must be also local in space. 
Thus, in the case when the generalized interaction 
operator $H^{(s)}_{int}(\tau)$ has the 
form (48), the theory is local in nature, and the introduction 
of a nonlocal form factor leads to the intrinsic inconsistency 
of the theory. 
On the other hand, our formalism permits a generalization 
to the class of the operators $H^{(s)}_{int}(\tau)$ that 
are not of the form (48). 
In this case the duration time $\tau = t_2 - t_1$ 
of the fundamental interaction is not zero, and 
the description of the dynamics of the system 
becomes fundamentally  nonlocal. 
For example, in QED the operator $H_{int}(t_2,t_1)$  
may be specified in the form 
$$
H_{int}(t_2,t_1) = \int d^3 x_1 \int d^3 x_2
{\cal H}_{int}(x_2,t_2,x_1,t_1),
$$
with 
$$
{\cal H}_{int}(x_2,x_1) = 
\int d^4 y_1 \int d^4 y_2
F(x_2-x_1, x_2-y_1, x_2-y_2) j_{\mu}(y_1) A^{\mu}(y_2).
$$
Here $F(x_2-x_1, x_2-y_1, x_2-y_2)$ is a relativistically 
invariant form factor. 
The matrix elements $<\psi_2|{\cal H}_{int}(x_2,x_1)|\psi_1>$
may be interpreted as amplitudes describing processes in which
interaction begins at a point $x_1$ of space-time and ends at a point
$x_2,$ and the presence of the nonlocal form factor in the
expression of ${\cal H}_{int}(x_2,x_1)$ seems to be natural.
One may hope that the form factor $F(x_2-x_1,x_2-y_1,x_2-y_2)$
can be chosen in such a way that no divergence difficulties
arise in describing the time evolution of systems in QED,
and, at the same time, solution of Eq.(37) leads to the ordinary
renormalized expression for the S matrix.

Let us now show that the proposed formalism may open new 
possibilities for describing the low-energy nucleon dynamics. 
For this purpose, come back to the model considered in Sec.VI. 
In the connection with the fact that this model 
is an extension of the ordinary separable-potential model 
to the case when the separable interaction is nonlocal
in both space and time,
note that separable potentials are 
widely used in nuclear physics (see, for example, Refs.[25-30]). 
One may hope that within this extended model 
a better agreement with experiment can be achieved. 
 
In the case when the operator 
$H_{int}(t_2,t_1)$ has such a form that the interaction 
generating the dynamics of a system in nonlocal 
in time, the effective interaction operator $B(z)$ 
is energy-dependent.
As has been mentioned above, the energy dependence of
the effective operators of hadron-hadron interactions
is associated with the quark and gluon degrees of freedom which are
not included explicitly in the description of low-energy
hadron interaction.
It should be noted that, in general, the energy-dependent 
operator of hadron-hadron interaction cannot be also 
assumed to be an optical potential, since due to the 
quark confinement a hadron system can be considered 
as a closed system, and for conservation of probability 
one need not to include explicitly in the description the 
channels associated with the quark and gluon degrees 
of freedom. 
A remarkable feature of the proposed formalism is that 
the operator of interaction may be energy-dependent 
even in the case of closed systems when the time 
evolution is unitary. 
At the same time, the quark and gluon degrees of freedom 
may manifest themselves in the dependence 
upon the time duration $\tau= t_2 -t_1$ 
of the interaction. 
In this case  the generalized interaction 
operator may be interpreted as an operator whose matrix 
elements $<\psi_2| H_{int}(t_2,t_1)|\psi_1>$
describe transitions between hadron states, and $t_1$
and $t_2$ are the times between  which the
quark and gluon degrees of freedom come into play. 
One may hope that it will become possible to construct 
such operators by using QCD - inspired quark models. 

Taking into account the quark and gluon degrees 
of freedom is especially important for describing the 
short-range NN interaction. 
In this connection note that in the most realistic NN 
potentials theoretical parts are supplemented with 
phenomenological short-range parts. 
Our formalism makes it possible to describe these 
phenomenological parts in terms of the operator 
$H_{int}(t_2,t_1).$ 
This may open new possibilities for reducing the still 
existing discrepancy  between  theory 
and experiment. 

\section*{Appendix A}

Represent (24) in the form 
$$
<n_2|R(t,t_0)|n_1> = \lim \limits_{a\to \infty}
<n_2|R_a(t,t_0)|n_1>, 
$$
where 
$$
<n_2|R_a(t,t_0)|n_1> = -\frac{a}{2\pi}  
\int^t_{t_0} dt_2 \int^{t_2}_{t_0} dt_1
\int^\infty_{-\infty} dx 
\frac{<n_2|T(z)|n_1>}{z+a} 
$$
$$
\times exp(iE_{n_2}t_2) 
exp[-iz(t_2-t_1)] exp(-iE_{n_1}t_1), 
$$
and $a$ is real.
Integrating over $t_1$ and $t_2$ (here 
the integrations over $t_1,t_2$ 
and $x_1$ may be reversed in order), we get  
$$
<n_2|R_a(t,t_0)|n_1> = \frac{a}{2\pi}  
\int^\infty_{-\infty} dx 
\frac{<n_2|T(z)|n_1>}{(z-E_{n_2})(z-E_{n_1})(z+a)} 
$$
$$
\times exp[-i(z-E_{n_2})t] 
exp[i(z-E_{n_1})t_0] +  
<n_2|B(t,t_0)|n_1>,  
$$
where 
$$
<n_2|B(t,t_0)|n_1> = \int^\infty_{-\infty} dx 
<n_2|F(z,t,t_0)|n_1>,  
$$
with 
$$
<n_2|F(z,t,t_0)|n_1> = -\frac{a}{2\pi}  
\frac{<n_2|T(z)|n_1>}{(z+a)} 
$$
$$
\times \left [ \frac {exp[i(E_{n_2} -E_{n_1})t_0]}
{(z-E_{n_2})(z-E_{n_1})} -  
\frac {exp[i(E_{n_2}-E_{n_1})t] - exp[i(E_{n_2}-E_{n_1})t_0]}
{(z-E_{n_2}) (E_{n_2}-E_{n_1})}
\right ] .  
$$
Taking into account the above analytic properties of 
$<n_2|T(z)|n_1>$ and the asymptotic condition (25), we get 
$$
\int^\infty_{-\infty} dx <n_2|F(z,t,t_0)|n_1> 
= \oint_{c_1} dz <n_2|F(z,t,t_0)|n_1> = 0, 
$$
where $ z=x+iy, y>0,$ and the contour of integration 
$c_1$ consists of the axis $y=constant$ and the upper half 
of the infinite circle. 
Thus for $<n_2|R(t,t_0)|n_1>$ we have 
$$
<n_2|R(t,t_0)|n_1> =
$$
$$=\lim \limits_{a \to \infty} \left (
\frac{a}{2\pi}  
\int^\infty_{-\infty} dx 
<n_2|T(z)|n_1>
 \frac {exp[-i(z-E_{n_2})t] exp[i(z-E_{n_1})t_0]}
{(z-E_{n_2})(z-E_{n_1})(z+a)} \right )  
$$
$$
= \frac{1}{2\pi}  
\int^\infty_{-\infty} dx 
 \frac {exp[-i(z-E_{n_2})t]exp[i(z-E_{n_1})t_0]}
{(z-E_{n_2})(z-E_{n_1})}
<n_2|T(z)|n_1> .  
$$

\section*{Appendix B}
Let us show that
the evolution operators given by (10) and (26) 
satisfies the composition law (4), 
provided equation (35) is valid. 
The right-hand side of (4) 
can be rewritten as follows:
$$
U_r(t,t^{\prime},t_0) \equiv
U(t,t^{\prime}) U(t^{\prime},t_0) =
{\bf{1}} + iR(t,t^{\prime})+ iR(t^{\prime},t_0) -
R(t,t^{\prime}) R(t^{\prime},t_0).
$$
Let $t \geq t^{\prime} \geq t_0.$ 
By using (26), for 
$<n_2| U_r(t,t^{\prime},t_0)|n_1>,$ 
we get
$$
<n_2|U_r(t,t^{\prime},t_0)|n_1> = <n_2|n_1> +
i <n_2|R(t,t^{\prime})|n_1> + 
$$
$$
+ i<n_2|R(t^{\prime},t_0)|n_1> +
<n_2| D(t,t^{\prime},t_0)|n_1>,
$$
where 
$$<n_2|D(t,t^{\prime},t_0)|n_1> = 
- \frac{1}{(2\pi)^2} 
\sum \limits_n \int dx_1 \int dx_2 
exp[-i(z_2-E_{n_2})t] \times
$$
$$
\times
exp[i(z_1-E_{n_1})t_0] exp[i(z_2-z_1)t^{\prime}]
 \frac {<n_2|T(z_2)|n><n|T(z_1)|n_1>}
{(z_2-E_{n_2})(z_2-E_n)(z_1-E_n)(z_1-E_{n_1})} ,
$$
with $z_1=x_1+ iy_1,$ $z_2=x_2+iy,$ $y_2 >y_1.$ 
Assuming that $T(z)$ satisfies equation (35),  this equation 
can be written in the form 
$$
<n_2|D(t,t^{\prime},t_0)|n_1> = 
- \frac{1}{(2\pi)^2} 
 \int dx_1 \int dx_2 \times
$$
$$
\times
\frac {exp[-i(z_2-E_{n_2})t]
exp[i(z_1-E_{n_1})t_0] exp[i(z_2-z_1)t^{\prime}]}
{(z_2-E_{n_2})(z_1-E_{n_1})}
$$
$$
\times \left \lbrack
\frac {<n_2|T(z_1)|n_1>}{z_2-z_1} - 
\frac {<n_2|T(z_2)|n_1>}{z_2-z_1} 
\right \rbrack  = \frac {i}{2\pi} \int dx_1 \times
$$
$$
\times
\frac {exp[-i(z_1-E_{n_2})t]
exp[i(z_1-E_{n_1})t_0]-
exp[-i(z_1-E_{n_2})t] exp[i(z_1-E_{n_1})t^{\prime}]}
{(z_1-E_{n_2})(z_1-E_{n_1})} \times 
$$
$$
\times
<n_2|T(z_1)|n_1> -
\frac {i}{2\pi} \int dx_2
\frac {exp[-i(z_2-E_{n_2})t^{\prime}]
exp[i(z_2-E_{n_1})t_0]}
{(z_2-E_{n_2})(z_2-E_{n_1})} <n_2|T(z_2)|n_1> =
$$
$$
= i (<n_2|R(t,t_0)|n_1> - 
<n_2|R(t,t^{\prime})|n_1> -
<n_2| R(t^{\prime},t_0)|n_1>).
$$
From this equation it  follows  that equation (4) is 
satisfied for $t \geq t^{\prime} \geq t_0.$
By using (13), it is easy to show then that the evolution 
operator satisfies the composition law (4) for any 
$t, t^{\prime}$ and $t_0.$

\section*{Appendix C}
Let us now derive an equation directly for the operator 
$\tilde S(t_2,t_1)$ starting from equation (35). 
From this equation, using the definition of $<n_2|T(z)
|n_1>$ given by (27), one can find
$$
\int^\infty_0 d\tau exp(iz_1\tau)\tau 
(1- exp[i(z_2 -z_1)\tau])
<n_2|\tilde T(\tau)|n_1> =
\sum_n \int^\infty_0 d\tau exp(iz_1\tau) 
$$
$$
\times 
\int^\tau_0 d\tau_2
\int_0^{\tau-\tau_2} d\tau_1
(\tau-\tau_1-\tau_2)
exp[-iE_n(\tau-\tau_1-\tau_2)]
<n_2|\tilde T(\tau_2)|n><n|\tilde T(\tau_1)|n_1> .
$$
For this relation to be valid for any $z_1$ and 
$z_2$ the matrix elements $<n_2|\tilde T(\tau)|n_1>$
must satisfy the following equation:
$$
(1- exp[i(z_2 -z_1)\tau])
\tau <n_2|\tilde T(\tau)|n_1> = \sum_n
\int^\tau_0 d\tau_2  \int_0^{\tau-\tau_2} d\tau_1 (\tau-\tau_1-\tau_2)
 $$
$$
\times exp[-iE_n(\tau-\tau_1-\tau_2)]<n_2|\tilde T(\tau_2)|n><n|
\tilde T(\tau_1)|n_1>.
$$
From this equation, by using (15), one can get the equation 
(36) for the operator $\tilde S(t_2,t_1)$. 
Equation (36)  has be rewritten in the form
$$
\left \lbrack 
\sum_{n=1}^{\infty} \frac {(ia)^n}{n!}(t_2-t_1)^n)
\right \rbrack
\tilde S(t_2,t_1) = \int^{t_2}_{t_1} dt_4
\int^{t_4}_{t_1}dt_3
$$
$$
\times 
\left \lbrack 
\sum_{m=0}^{\infty} \frac {(ia)^m}{m!}(t_2-t_4)^m
\right \rbrack
\left \lbrack 
\sum_{k=1}^{\infty} \frac {(ia)^k}{k!}
(t_4-t_3)^k
\right \rbrack
\tilde S(t_2,t_4) \tilde S(t_3,t_1).
$$
For this relation to be valid for every $a$, 
each term of the 
expansion in powers of $a$ on the left-hand side 
of this equation must be equal to the term of the same 
order on its right-hand side. 
In the first order, for example, we have 
$$
a(t_2-t_1)\tilde S(t_2,t_1) = 
\int^{t_2}_{t_1} dt_4 \int^{t_4}_{t_1}dt_3
(t_4-t_3) \tilde S(t_2,t_4) \tilde S(t_3,t_1).
$$
From this it follows that $\tilde S(t_2,t_1)$ 
must satisfy the equation (37). 

\section*{Appendix D}

Let us  show that in the case  $\alpha <1/2$ 
there are states in the Hilbert space for which 
$<\psi_2|U(t,0)|\psi_1>$ 
are not continuous. 
Using (10), (12) and (15), for $<\psi_2|R(t,0)|\psi_1>,$ 
we can write 
$$
<\psi_2|R(t,0)|\psi_1> = 
i \int_0^t dt_2 \int_0^{t_2} dt_1 \int d^3 k_1 \int d^3 k_2
exp[i(E_{k_2}-E_{k_1})t_2] \times
$$
$$
\times
exp[iE_{k_1}(t_2-t_1)] 
\psi_2^*({\bf k}_2) \psi_1({\bf k}_1) 
<{\bf k}_2|\tilde T(t_2-t_1)|{\bf k}_1>,
$$
where 
$\psi_i({\bf k}) = <{\bf k}| \psi_i>, i=1,2.$
Consider the vectors $|\psi_{\nu}>$ 
for which 
$$
<{\bf k}| \psi_{\nu}> =  
\frac 
{\nu^{\frac{1}{2}} d}
{k(k-\nu-i\nu \Gamma_0)},
$$
where  $d= \Gamma_0^{1/2} (2\pi)^{-1/2} [\pi/2 +
arctg(\frac {1}{\Gamma_0})^{-1}], $
$k=|{\bf k}|,$ 
and $\Gamma_0$ is a real constant.
It is easy to verify that 
these vectors are normolized.  

Let us now consider the case  $\alpha >1/2.$
Taking into account (56) and (63), for 
$<\psi_{\nu}|R(t,0)|\psi_{\nu}>$ in the limit 
$\nu \to \infty,$ we have 
$$
<\psi_{\nu}|R(t,0)|\psi_{\nu}> = 
\nu^{1-2\alpha} gd^2 |c_1|^2
\int_0^{\theta} d\theta_1 
\int d^3 q_1 \int d^3 q_2
\times
$$
$$
\times
\frac {exp[i(E_{q_2}- E_{q_1}) \theta_1]}
{(q_1- 1+i\Gamma_0)(q_2-1-i\Gamma_0) 
q_1^{1+\alpha}q_2^{1+\alpha}}
+ o(\nu^{1-2\alpha}), 
$$
where $q_i=k_i \nu^{-1}, \theta=t \nu^2, 
\theta_1 = t_1 \nu^2.$
It is easy to see that 
$<\psi_{\nu}|R(t,0)|\psi_{\nu}>$ 
given by this equation tends to zero as $\nu \to \infty.$
Let us consider now the case  $\alpha<1/2.$
According to (69),  
$<{\bf k}_2 |\tilde T(\tau)|{\bf k}_1>$ 
has the following behaviour near 
the point $\tau=0:$
$$
<{\bf k}_2 |\tilde T(\tau)|{\bf k}_1> =
a_1 \varphi({\bf k}_2) \varphi^*({\bf k}_1) \tau^{-\beta}
+ o(\tau^{-\beta}),
$$
with $\beta=  \alpha+1/2.$
Taking into account this fact, for 
$<\psi_{\nu}|R(t,0)|\psi_{\nu}>$ in the limit 
$\nu \to \infty,$ we get 
$$
<\psi_{k_0}|R(t,0)|\psi_{k_0}> = 
\nu^{2\beta-1-2\alpha} \int_0^{\theta} d \theta_2
\int_0^{\theta_2} d\theta_1 F_2(\theta_1,\theta_2) 
+ o(\nu^{\alpha+ 1/2 - \beta}), 
$$
where 
$$
F_2(\theta_1,\theta_2) =
a_1 d^2 |c_1|^2
\int d^3 q_1 \int d^3 q_2 
\frac {exp[i(E_{q_2}- E_{q_1}) \theta_2] 
exp[iE_{q_1} (\theta_2 - \theta_1)]}
{(q_1- 1+i\Gamma_0)(q_2-1-i\Gamma_0) 
q_1^{1+\alpha}q_2^{1+\alpha} (\theta_2-\theta_1)^{\beta}}
, $$
${\bf q}_i={\bf k}_i\nu^{-1}, \theta=t_i \nu^2,$ and  
$\theta = t \nu^2.$
Since $\beta= \alpha+1/2,$  it follows from this that 
$$
\lim \limits_{\nu \to \infty}
<\psi_{\nu}|R(t,0)|\psi_{\nu}> = C,
$$
where 
$ C=  \int_0^{\infty} d \theta_2
\int_0^{\infty} d\theta_1 F_2(\theta_1,\theta_2) .$
Thus, in the limit $\nu\to \infty$ the matrix 
element $<\psi_{\nu}|R(t,0)|\psi_{\nu}>$ 
is independent of $t.$ 
This means that there are normalized vectors in the 
Hilbert space for which 
$<\psi_2|U(t,0)|\psi_1>$ does not tend to $<\psi_2|\psi_1>$ 
as $t \to 0.$ 
However, these vectors 
represent the states with infinite momentum and correspondingly 
with infinite energy which are not physically realisable 
and hence the condition (7) is not violated.

\newpage
\section*{References}

\begin{enumerate}

\item[{[1]}]
  N.N. Bogolyubov and D.V. Shirkov,
 Introduction to the Theory of Quantized Fields, 3rd.ed.
(Wiley, New York, 1979).
\item[{[2]}]
M.A. Braun,  Zh. Eksp. Teor.Fiz.{\bf 94}, 145 (1988)
[Sov.Phys. JETP {\bf 67}, 2039 (1988)].
\item[{[3]}]
Yu.S. Kalashnikova, I.M.Narodetsky, and V.P.Yurov,  
Yad. Fiz., {\bf 49}, 632 (1989).
\item[{[4]}]
 A.G. Baryshnikov, L.D.Blokhintsev, I.M.Narodetsky, and D.A.Savin, 
 Yad. Fiz. {\bf 48}, 1273(1988).
\item[{[5]}]
 A.N. Safronov, Teor.Mat.Fiz. {\bf 89}, 420(1991); 
 Yad.Fiz., {\bf 57}, 208(1994).
\item[{[6]}]
 Yu.A. Kuperin, K.A. Makarov, and S.P.Merkuriev, 
Teor.Mat.Fiz., {\bf 75}, 431(1988);
{\bf 76}, 242(1989).
\item[{[7]}]
 A. Abdurakhmanov and A.L. Zubarev, Z.Phys. A {\bf 322}, 523(1985).
\item[{[8]}]
 M. Orlowski,  Helv.Phys.Acta. {\bf 56}, 1053(1983).
\item[{[9]}]
 B.O. Kerbikov, Yad.Fiz. {\bf 41}, 725(1985);
 Teor. Mat.Fiz. {\bf 65}, 379(1985).
\item[{[10]}]
 Ya.A.Simonov,  Phys.Lett. B, {\bf 107}, 1(1981). 
\item[{[11]}]
 C.R. Chen, G.L.Payne, J.L.Friar, and B.F.Gibson, 
 Phys. Rev. {\bf 31}, 2266(1985).
\item[{[12]}] 
R.P. Feynman, Rev. Mod. Phys. {\bf 20}, 367(1948).
\item[{[13]}]
  R.P. Feynman  and A.R. Hibbs, Quantum Mechanics and Path
  Integrals  (McGraw-Hill, New York, 1965).
\item[{[14]}] 
 R.F.Streater and A.S.Wightman, PCT, 
 Spin and Statistics, And All That.
 (W.A.Benjamin, New York, 1964). 
\item[{[15]}] 
 N.N.Bogolyubov, B.V.Medvedev, and M.K.Polivanov, 
 Theory of Dispersion Relations. 
 (Lawrence Rad. Lab., Berkeley, Calif., 1961).
\item[{[16]}]
M. Reed and B. Simon,
Methods of Modern Mathematical Physics I 
(Academic Press, New York, 1972).
\item[{[17]}]  
R.Kh. Gainutdinov, Yad. Fiz. {\bf 37}, 464 (1983)
 [Sov.J.Nucl.Phys. {\bf 37}, 277(1983)].
\item[{[18]}]  
R.Kh. Gainutdinov,  Yad. Fiz. {\bf 46}, 1271(1987)
 [Sov.J.Nucl.Phys. {\bf 46}, 743(1987)].
\item[{[19]}]  
R.Kh. Gainutdinov,  J. Phys. A. {\bf 22}, 269(1989).
\item[{[20]}]  
R.Kh. Gainutdinov,  Yad. Fiz. {\bf 53}, 1431(1991) 
[Sov.J.Nucl.Phys. {\bf 53}, 885(1991)].
\item[{[21]}]
  R.Kh. Gainutdinov, Zh. Eksp. Teor. Fiz. {\bf 108}, 1600 (1995)
 [Sov. Phys. JETP {\bf 81}, 877(1995)].
\item[{[22]}]  
R. Kh. Gainutdinov and  A.A.Mutygullina, 
 Yad. Fiz. {\bf 60}, 938 (1997) [Physics of Atomic Nuclei, 
 {\bf 60}, 841 (1997)].
\item[{[23]}] 
R.Kosloff, Ann.Rev.Phys.Chem., {\bf 45}, 145 (1994).
\item[{[24]}] 
H.J.L{\"u}dde, A.Henne, T.Kirchner and R.M.Dreizler,
J.Phys.B: At.Mol.Opt.Phys. {\bf 29}, 4423 (1996).
\item[{[25]}] 
Y.Yamaguchi,  Phys.Rev. {\bf 95}, 1635 (1954).  
\item[{[24]}]  
F.Tabakin,  Phys.Rev.  {\bf 174}, 1208 (1968).
\item[{[27]}] 
J. Haidenbauer and W. Plessas, Phys.Rev. C {\bf 30}, 1822 (1984). 
\item[{[28]}]  
G.Rupp and J.A.Tjon, Phys.Rev.C  {\bf 37}, 1729 (1988).
\item[{[29]}]  
T. Alm and G.R{\"o}pke, Phys.Rev.C  {\bf 50}, 31 (1994).
\item[{[30]}]  
S.K.Adhikari  and L.Tomio, Phys.Rev.C  {\bf 51}, 70 (1995).

\end{enumerate}
\end{document}